\documentclass[useAMS,usegraphicx]{mn2e}
\usepackage{rotate}
\usepackage{times}
\newif\ifAMStwofonts
\AMStwofontstrue

\def\gsim{\mathrel{\hbox{\rlap{\hbox{\lower4pt\hbox{$\sim$}}}\hbox{$>$}}}}
\def\lsim{\mathrel{\hbox{\rlap{\hbox{\lower4pt\hbox{$\sim$}}}\hbox{$<$}}}}

\def\fxfopt{{$F_{\rm X}/F_{\rm opt}$}}

\title{X-ray and optical counterparts of hard X-ray selected sources from
the SHEEP survey: first results}
\author[K. Nandra et al.]{K. Nandra$^{1,2}$, I. Georgantopoulos$^{3}$,
M. Brotherton$^{4,5}$, I.E. Papadakis$^{6,7}$ \\ 
$^1$Astrophysics Group, Imperial College London, Blackett Laboratory,
Prince Consort Road, London SW7 2AW, UK \\ 
 $^2$Visiting Astronomer, Kitt Peak National Observatory \\
$^3$Institute of Astronomy \& Astrophysics, National Observatory of
Athens, I. Metaxa B. Pavlou, Penteli, 15236, Athens, Greece \\
$^4$Department of Physics and Astronomy, University of Wyoming,
Laramie, WY 82071, USA \\
$^5$Kitt Peak National Observatory \\
$^6$Department of Physics, University of Crete, 71 003, Heraklion,
Greece \\
$^7$IESL, FORTH-Hellas, 71 110,Heraklion, Crete, Greece 
}
\date{}

\begin{document}

\maketitle
\label{firstpage}

\begin{abstract}
We present followup observations of five 
hard X-ray sources
from the ASCA 5-10 keV SHEEP survey, which has a limiting
flux of $\sim 10^{-13}$ erg cm$^{-2}$ s$^{-1}$. 
Chandra data have been obtained to improve the X-ray positions
from a few arcmin to $<1''$, which allows unambiguous optical
identification. While the objects almost certainly house AGN
based on their X-ray luminosity, optical spectroscopy
reveals a variety of properties.
The identifications indicate that the SHEEP survey samples the same
populations as deeper surveys which probe the origin of the X-ray background,
but because the SHEEP sources are far brighter, they are more
amenable to detailed followup work. We find a variety of classifications and properties,
including a type II QSO, a galaxy undergoing star formation, 
and a broad-line AGN  which has a very hard X-ray spectrum, indicating substantial absorption in the X-ray but none in the optical.
Two objects have X-ray/optical flux ratios which, were they
at an X-ray flux level typical of objects in Chandra deep surveys, 
would place them in the ``optically faint'' category.
They are both identified with broad line QSOs at z$\sim 1$.
Clearly this survey - which is relatively unbiased against obscured objects
- is revealing a set of 
remarkable objects quite different to the familiar classes of AGN found in previous
optical and soft X-ray surveys.
 \end{abstract}


\section{INTRODUCTION}
\label{Sec:Introduction}

The majority of the X-ray background in the 2-8 keV band has been
resolved into discrete sources, thanks mainly to Chandra deep surveys
(Mushotzky et al. 2000; Brandt et al. 2001; Giacconi et al. 2001;
Cowie et al. 2002). The majority of the objects are almost certainly
AGN, and if so they are more numerous than active galaxies found
by other surveys. These newly discovered objects,
many of which have hard X-ray spectra, are therefore the dominant AGN
population and our understanding of black hole accretion, and its
evolution through cosmic time, depends on their detailed
study. Mutiwavelength followup of these objects (e.g. Mushotzky et al
2000; Barger et al. 2001, 2002; Rosati et al. 2002) has shown that
they would have been very difficult to identify as AGN without the
X-ray observations. The reasons are twofold: 1) Spectroscopic followup
in the optical shows many objects have little or no evidence for the
high excitation and/or broad emission lines hitherto considered the
defining characteristic of AGN and 2) Many are too faint optically for
spectroscopic followup at all. Nonetheless, the X-ray luminosity of these objects,
typically $L_{\rm X} \sim 10^{43}$~erg s$^{-1}$ (Cowie et al. 2003), is strongly
indicative of AGN activity. 

The precise nature of the X-ray populations are therefore in doubt, but the best
bet is that they are obscured AGN at moderate redshift (Alexander et
al. 2001). Definitive answers will not be obtained unless they are
observed spectroscopically with larger telescopes, or we can select
brighter examples for study with the current generation of
ground-based facilities. The most
promising way currently is to investigate the bright end of the new X-ray
populations is by using large area surveys such as the BeppoSAX HELLAS survey
(Fiore et al. 1999; 2001; Comastri et al. 2001)
and our own equivalent  with ASCA, SHEEP (the Search for the High Energy
Extragalactic Population; Nandra
et al. 2003, hereafter paper I). 
The hard detection bandpass (5-10 keV) is relatively unbiased to absorption,
and the wide area means rarer, bright objects are found.

The big disadvantage of the HELLAS and SHEEP survey data is the
relatively poor angular resolution of the BeppoSAX and ASCA
telescopes, which make it very difficult to identify the X-ray sources
unambiguously. To improve on this, we are obtaining Chandra data for a
subset of the SHEEP objects, which should give arcsec position and
therefore secure optical counterparts for optical spectroscopic and other 
followup. The first results from this program are the
subject of this Letter.

\section{OBSERVATIONS AND DATA ANALYSIS}

The SHEEP survey comprises 69 objects (Paper I). 35 of these have been
detected by ROSAT, and of these 13 have reasonably secure optical
counterparts in archival catalogs. For the remainder of the ROSAT detected
objects, we are in the process of obtaining optical imaging and
spectroscopy of the likely counterparts, with full details presented
elsewhere (Zezas et al., in preparation). For the remaining objects either
unobserved or undetected by ROSAT, we are obtaining Chandra data to
provide subarcsec positions for unambiguous optical identification.
Here we present a subset of 5 objects for which we have now obtained 
spectroscopic classifications. No other selection was
applied in choosing these sources other than the fact that
they have new optical identifications, 

\begin{table*}
\centering
\caption{X-ray data for the sample.
Col.(1): ASCA designation;
Col.(2): Chandra sequence number
Col.(3): Exposure time
Col.(4) Chandra X-ray counterpart, or ROSAT (RX) for S5;
Col.(5): Chandra counts in 2-10 keV band;
Col.(6): Observed frame 2-10 keV flux in units of $10^{-13}$~erg cm$^{-2}$ s$^{-1}$ converted assuming a $\Gamma=1.6$ spectrum ($2.43 \times 10^{-11}$~erg cm$^{-2}$ ct$^{-1}$ from
Chandra or ASCA (S5 only);
Col.(7): log of 2-10 keV rest frame luminosity in erg s$^{-1}$ for $h=0.7$, $\Omega_{\Lambda}=0.7$, $\Omega_{M}=0.3$;
Col.(8): Hardness ratio (H-S)/(H+S) where H=counts in 2-10 keV band
and S=counts in 0.5-2 keV band from Chandra (where available); Col(9): Optical R
magnitude; Col(10) Optical B magnitude; Col(11) Redshift; Col (12) Optical classification
\label{tab:sample}}
\begin{center}
\begin{tabular}{cccccccccccc}
\hline
Object Name & Chandra & Exp. & X-ray ID & Cts & $F_{\rm X}$ & $\log L_{\rm X}$ & HR1 & R & B & z & ID \\
(AX) & Sequence & (s) & (CXOU) &  &  2-10 keV & 2-10 keV &  Ratio & mag & mag & & \\
(1) & (2) & (3) & (4) & (5) & (6) & (7) & (8) & (9) & (10) & (11) & (12) \\
\hline
J$0140.1+0628$   (S1) & 900149 & 5283 & J014010.2+062827 & 43 & $1.98 \pm 0.30$ & 44.2 & 
+0.72  & 18.9 & 19.7 & 0.50 & QSO I \\
J$0836.2+5538$  (S2)  & 900155 & 4392 & J083622.8+553853 & 24 & $1.33 \pm 0.27$ & 45.0 &
-0.49 & 20.2 & 21.4 & 1.29 & QSO I \\
J$0836.6+5529$   (S3) & 900156 & 4202 & J083632.9+552847 &  4   & $0.23\pm 0.12$ & 44.0 & 
-0.80  & 20.8 & 21.9 & 0.98? & QSO I \\
J$1035.1+3938$   (S4) & 900158 & 5055 & J103515.6+393909 & 28 & $1.35 \pm 0.19$ & 42.6 &
0.75 & 17.2 & 18.6 & 0.11 & H{\sc ii} \\
J$1153.7+4619$   (S5) & ... & ... & J115345.6+462022 & ... & $5.20 \pm 1.34$ & 44.8 & ... & 
20.6 & $>$22.9 & 0.59 & QSO II \\
\hline
\end{tabular}
\end{center}
\end{table*}

\subsection{X-ray data}

\begin{figure}
\scalebox{0.5}{{\includegraphics{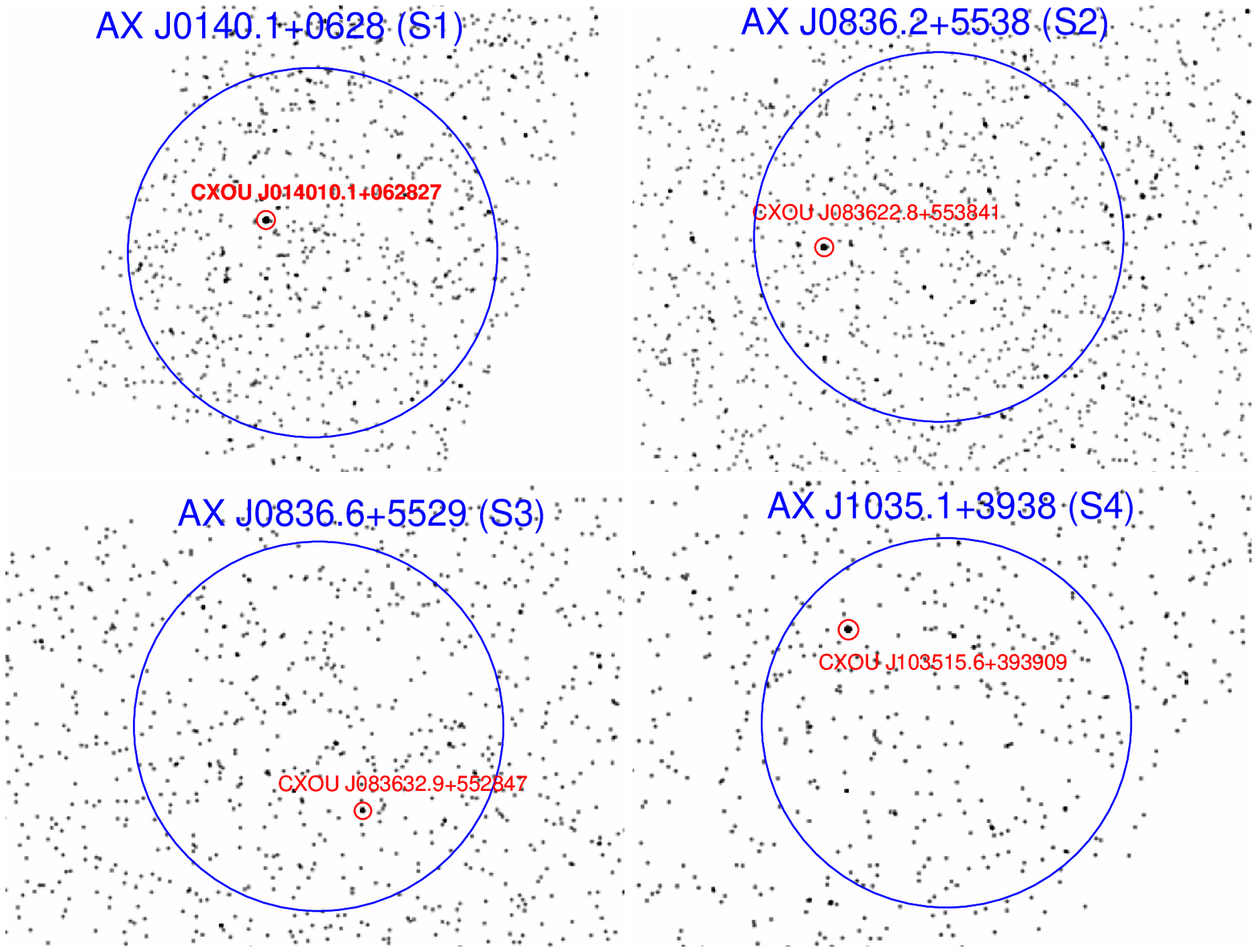}}}
\caption{Chandra 2-10 keV images of the SHEEP targets. The images are approximately $6^{\prime}.5$ square with N upwards and E to the left. The brightest
hard X-ray source detected by Chandra, which is the probable
counterpart of the SHEEP object, is marked by the small circle. In one case,
AX J0836.2+5538 (S2), two significant
sources are detected in the ASCA error circle (larger circles,
$2^{\prime}$ radius). The marked object is a factor
$\sim 6$ brighter, however, and is almost certainly the SHEEP
source. \label{fig:xim}}
\end{figure}

Details of the X-ray data for the sample are given in Table~\ref{tab:sample}. 
Henceforth
we refer to the sources as S1-S5 designated in the Table. 
The objects in the SHEEP survey were initially identified from ASCA
GIS images in the 5-10 keV band (Paper 1). It is very difficult to
identify an unambiguous optical counterpart from the GIS data alone,
but ROSAT positions were available for some of the objects, allowing
more secure identification. This was the case for only one of the 5
objects we discuss here, S5.  For the others, we have
obtained Chandra images.  The Chandra data were all obtained with the
ACIS-S3 back illuminated chip, and had nominal exposures of 5ks.
Observation details are shown in Table~\ref{tab:sample}. The data were
processed using the standard CXC software and we created images in multiple
bands from the level 2 event file. The 2-10 keV images are shown in Fig.~\ref{fig:xim}. We
performed point-source searches in the 5-10 keV, 2-10 keV and 0.5-2
keV band on the images using the Chandra wavdetect algorithm, with
a significance threshold of $10^{-6}$. In two of
the four cases (S1 and S4) the Chandra counterpart was very clear, with a single,
bright source being detected in the 2-10 keV band within $2^{\prime}$
radius (the approximate GIS error circle) of the nominal ASCA position. In
these two cases, the same object was also uniquely detected in the 5-10 keV band.

For S2, two objects are detected 
in the error circle at 2-10 keV, but one is  $\sim 6$ times brighter and is the only one
detected at 5-10 keV.  We therefore associate this brighter object with the SHEEP source. 
For S3 one object was detected at 2-10 keV, but none was
detected in the 5-10 keV image.
We assign the 2-10 keV source as the counterpart but note that this object has a
flux well below the original ASCA detection threshold. We discuss
this further below, but note 
that the lack of a clear bright, hard
counterpart does leave some ambiguity about the identification in this
case.

\subsection{Optical data}

All targets were imaged from the 1.3 m, f/7.7 Ritchey-Cretien
telescope at Skinakas Observatory in Crete, Greece, on November 14,
2001, except for S4 which was observed on June 3, 2002
and S5 which was observed on June 1, 2000.
The observations were carried out through the standard Johnson $B$ and
Cousins $R$ filters. The CCD used was a $1024 \times 1024$ SITe chip
with a 24 $\mu$m$^{2}$ pixel size (corresponding to
$0^{\prime\prime}.5$ on sky). The exposure time was 20 min for both
filters. The observations were done under photometric conditions, and
the seeing was between $\sim 1^{\prime\prime} - 2^{\prime\prime}$.
Standard image processing (bias subtraction and flat fielding using
twilight-sky exposures) was applied to all frames. In all cases, an
optical source was detected within $1^{\prime\prime}$ of the Chandra
source. Only S4 was resolved. For the others
we performed aperture photometry of the sources by integrating
sky-subtracted counts within a circular aperture of radius equal to 4 times
the seeing full-width at half maximum of each frame. For
S4, the extension is at the 10-12'' level and we extracted the counts 
from a 10'' radius aperture. 
Instrumental magnitudes were
transformed to the standard system through observations of more than
40 standard stars from Landolt (1992) during both nights. The B and R magnitudes of the objects are given in Table~\ref{tab:sample}. Typical photometric errors are 0.2 mag in B and 0.1 mag in R. S5 was not detected in B and the 5$\sigma$ upper limit is given. 

\begin{figure}
\rotatebox{270}
{\scalebox{0.2}{
{\includegraphics{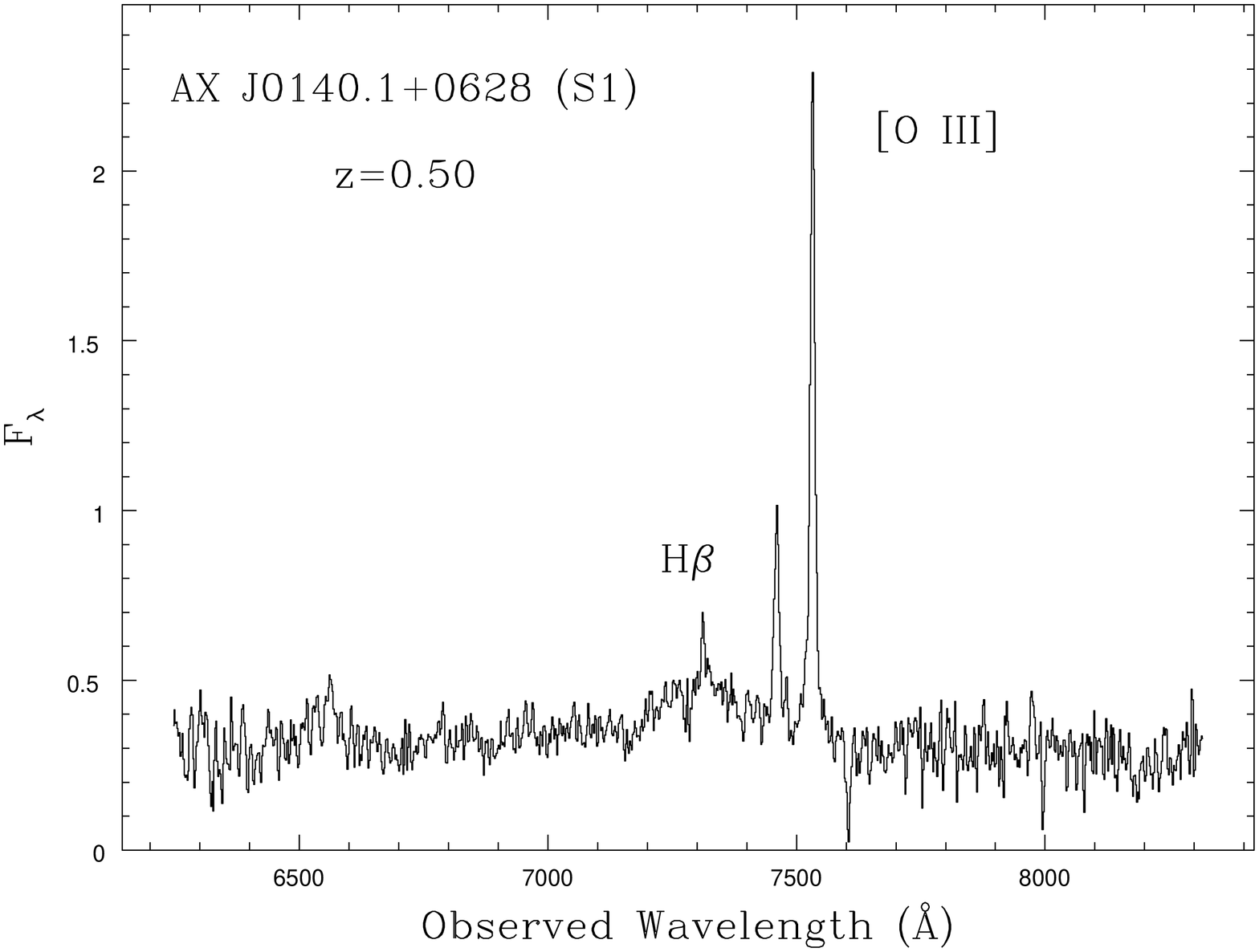}}
{\includegraphics{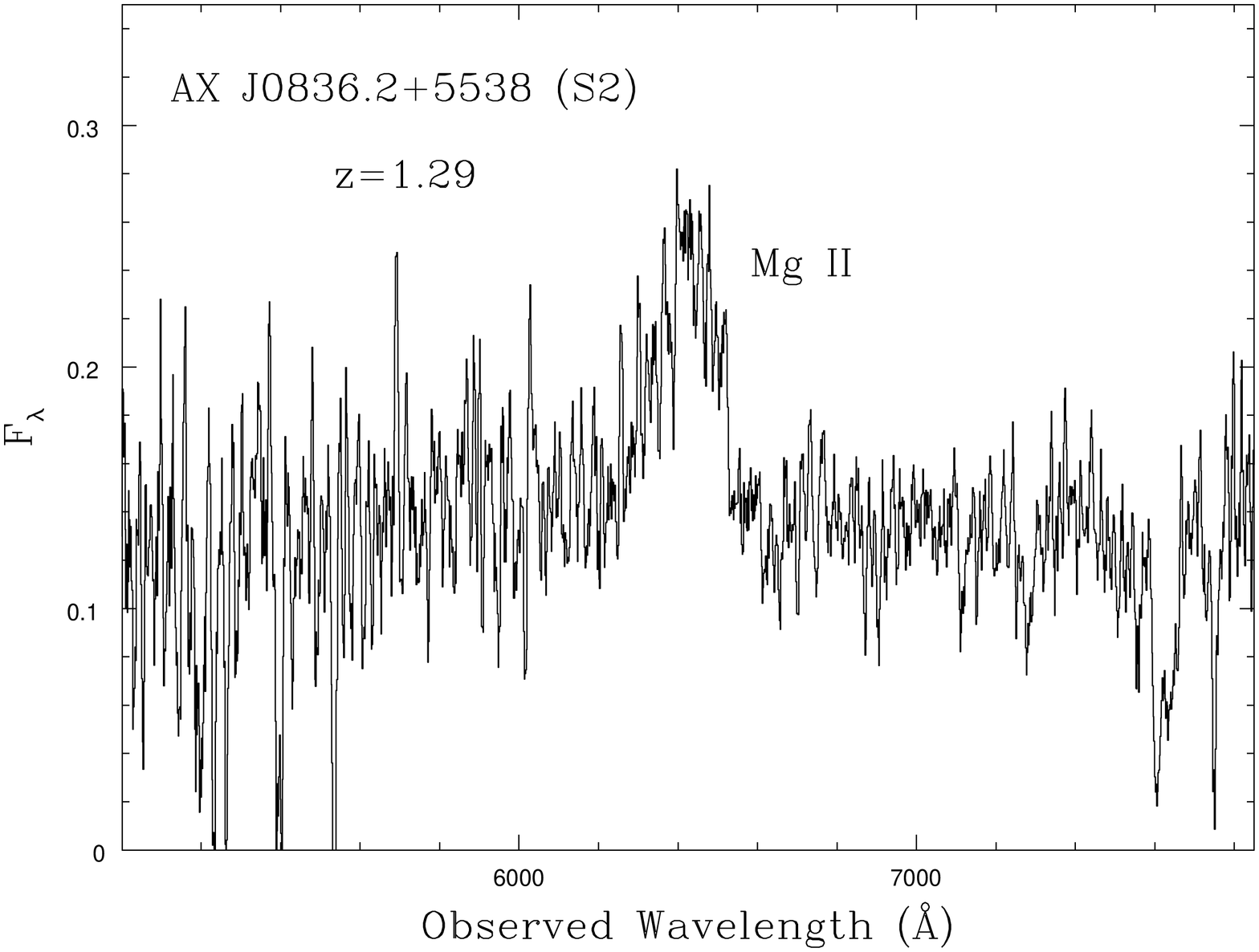}}
{\includegraphics{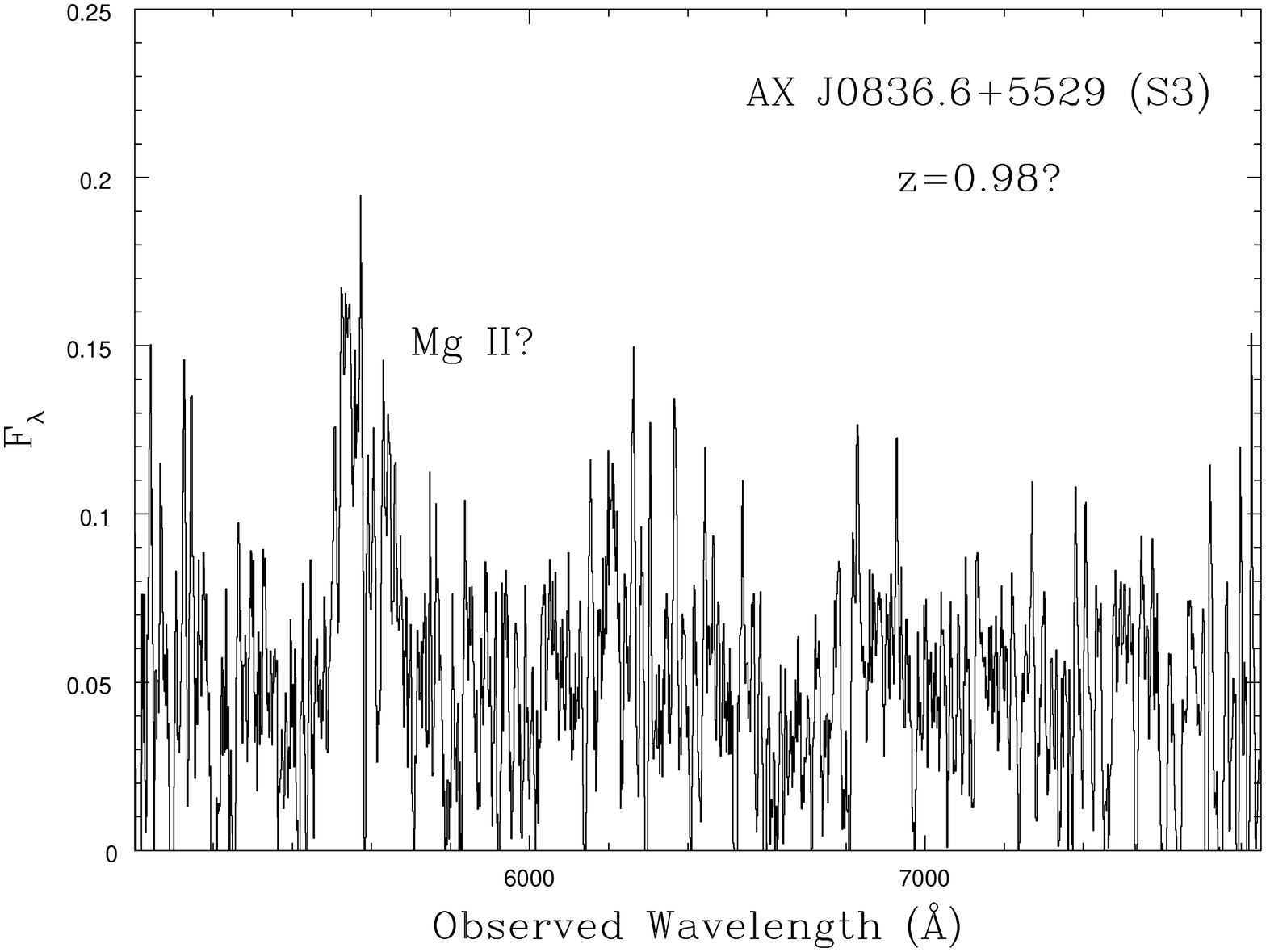}}
{\includegraphics{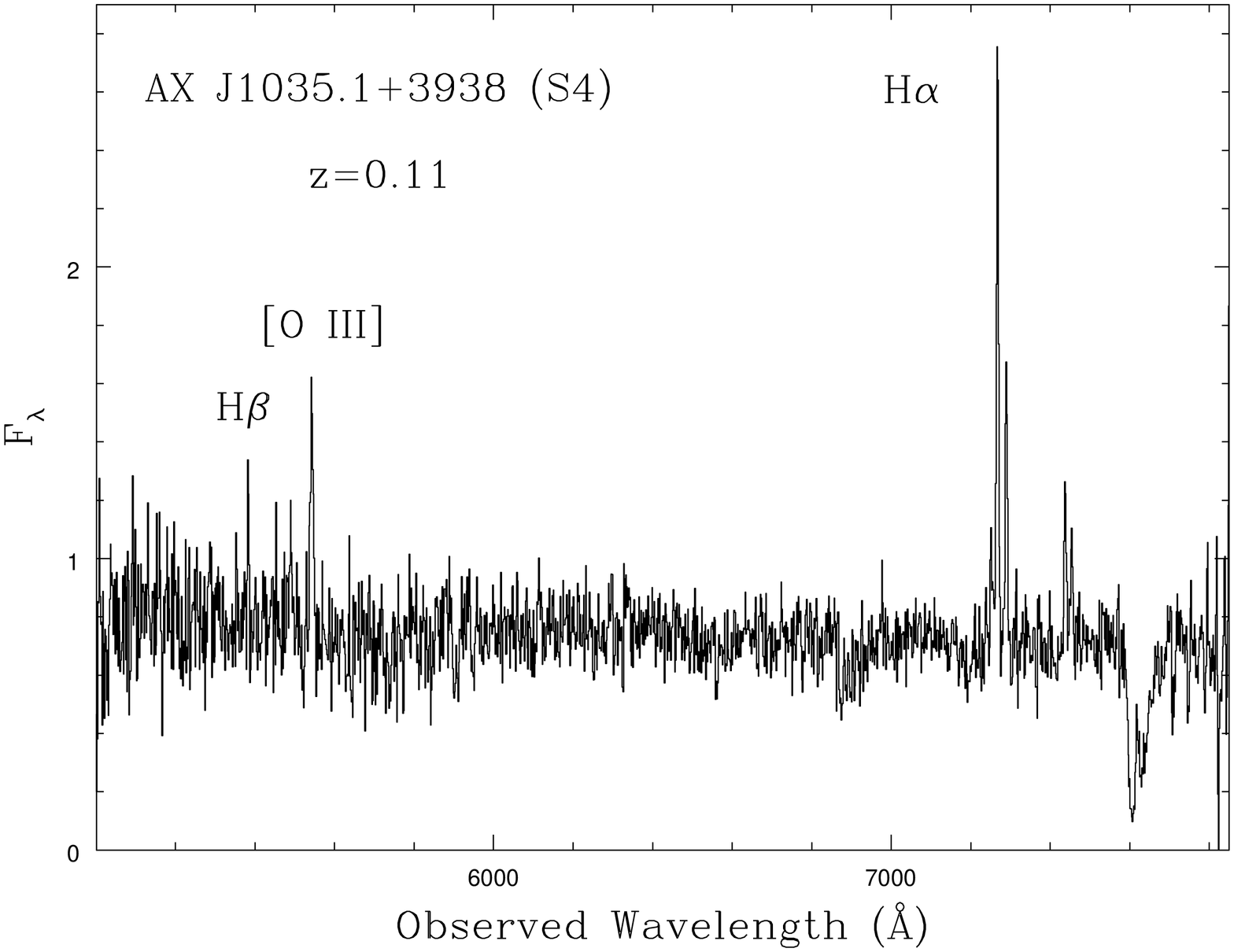}}
{\includegraphics{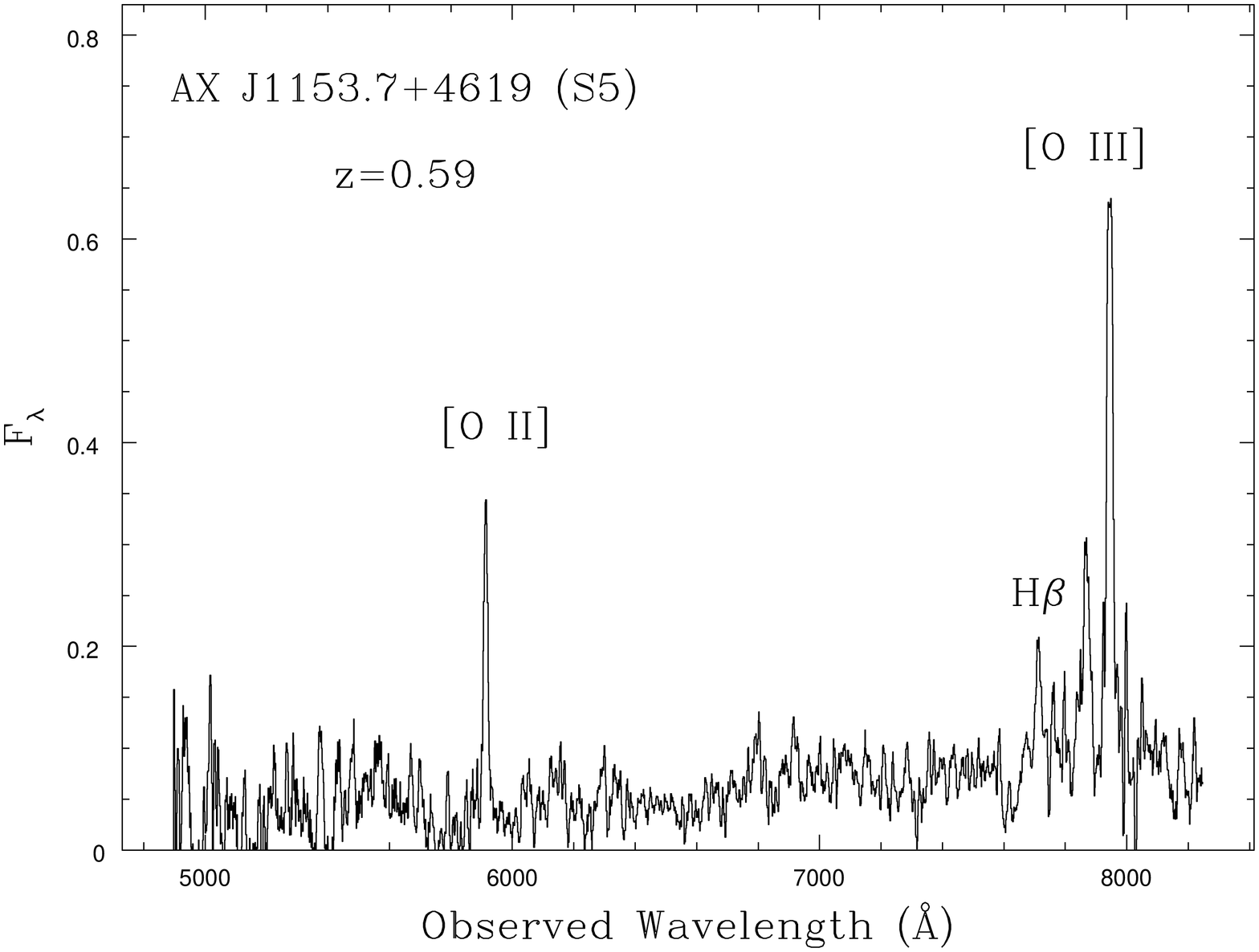}}
}}
\caption{Optical spectra for the 5 SHEEP objects. They were obtained with the RCSP at the KPNO 4-m telescope, with the exception of AX J0140.1+00628 (S1) which was taken with the  modular spectrograph at the MDM observatory. The fluxes are in units of $10^{-16}$ erg cm$^{-2}$ 
\AA$^{-1}$. Prominent emission lines are marked. The spectra of AX J0836.2+5538 (S2) and
AX J0836.6+5529 (S3) have been smoothed for display purposes. \label{fig:optspec}}
\end{figure}

We obtained optical long-slit spectroscopy using the Ritchey-Chretien
Focus Spectrograph (RCSP) mounted on the 4-m Mayall telescope at
Kitt Peak National Observatory (KPNO) on the nights of 17, 18 May 2002
(UT).  We used a new CCD detector, manufactured at Lawrence Berkeley
National Laboratory and designated LB1A, with superior red 
sensitivity in conjunction with a 316 l/mm grating blazed at 7500 \AA\
(BL181), an order-blocking filter, and a slit width of 1.5 arcsec
to provide coverage from
 $\sim$5000 to nearly 8000 \AA\ at a
resolution of $\sim$300 km s$^{-1}$. Conditions were photometric
with seeing varying from 0.7 arcsec to 1.3 arcsec. Exposure times
ranged from 40 minutes to two hours.  Multiple exposures were
weighted by their counts when combining to create final spectra. We
employed standard data reduction techniques within the NOAO IRAF
package and show the final spectra in Fig.~\ref{fig:optspec}.
The data for S1 were obtained by J. Halpern (priv. comm) at the 2.4m telescope of the MDM Observatory at KPNO in 1998 Dec 20. A spectrum with a resolution of 4\AA\ in the 6300-8300\AA\ range was obtained with the Modular Spectrograph in a 3600s exposure. This spectrum is also shown in Fig.~\ref{fig:optspec}. 

\section{RESULTS}

The Chandra observations have picked up an unambiguous  X-ray counterpart to the SHEEP source in three of the four cases. Typical fluxes in the 2-10 keV band are $1-2 \times 10^{-13}$~erg cm$^{-2}$ s$^{-1}$ (Table~\ref{tab:sample}). For S3, however, the likely Chandra counterpart has a flux far below the detection threshold of the SHEEP survey (paper I). This could be explained by Eddington bias in the original SHEEP survey (the initial ASCA detection was at only 4.9$\sigma$) or variability. Most likely, however, is a combination of both. If the main class of objects detected in any particular survey is variable, that survey will tend to pick out objects undergoing a (randomly) positive variability fluctuation, as well as a randomly high Poisson noise fluctuation. Indeed it is noteworthy that all the Chandra objects have 2-10 keV fluxes below that detected by ASCA. The Chandra observations do reveal objects with very hard X-ray spectra. This is demonstrated in Table~\ref{tab:sample}, which shows the hardness ratio derived
from the
2-10 keV and 0.5-2 keV fluxes (designated HR1 in Paper I). Aside from S3, the objects show hardness ratio inconsistent with an unabsorbed, $\Gamma=2$ spectrum typical of, e.g., soft X-ray selected QSOs (Georgantopoulos et al. 1997; Blair et al. 2000). 

\begin{figure}
\rotatebox{270}
{\scalebox{0.35}{
{\includegraphics{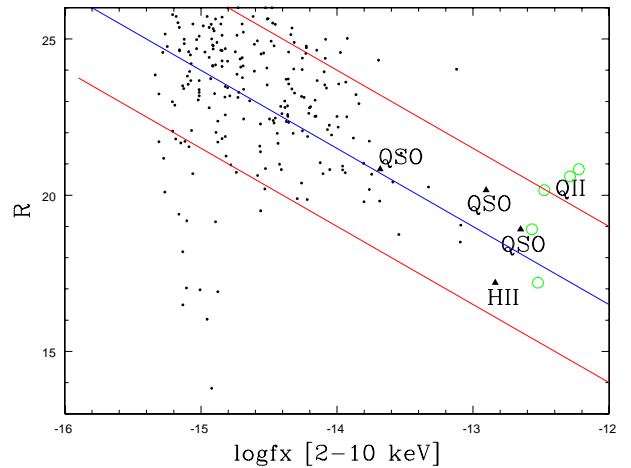}}
}}
\caption{R-band magnitude vs. 2-10 X-ray flux for the SHEEP objects (solid triangles: Chandra flux; open circles: ASCA flux. ASCA flux only for S5, the QII). 
The small dots are objects from the CDF-S survey (Giacconi et al. 2002).
The solid lines correspond to $\log$  \fxfopt = -1, 0 and 1, which encompasses most AGN. Our objects tend to show high \fxfopt\ ratio, with the exception of the HII galaxy. The ASCA \fxfopt\ ratios are particularly extreme in three cases, exceeding the $\log$  \fxfopt = 1 line, but these may be biased somewhat by selection effects, Eddington bias, and variability (see text). 
 \label{fig:fxfopt}}
\end{figure}

As we have already mentioned, the precise Chandra positions have also allowed us to assign an unambiguous optical counterpart to the X-ray sources. We plot in Fig.~\ref{fig:fxfopt} the 2-10 keV flux versus the $R$ magnitude determined from the Skinakas imaging. We have plotted both the ASCA flux (determined from the 5-10 keV count rate converted with a $\Gamma=1.6$ spectrum) and the Chandra flux (from the 2-10 keV count rate and the same spectrum). This illustrates the discrepancy between the fluxes noted above. In addition it shows where the objects lie in relation to the typical \fxfopt\ ratio for AGN, and in comparison to objects from the Chandra deep survey of the CDF-S (Giacconi et al. 2002).  With the position and magnitude information for all 5 targets, we have been able to identify all of them optically. We now discuss these identifications and the properties of the individual objects.

{\bf AX J0140.1+0629 (S1):} Chandra reveals an object with a very hard
spectrum, with 43 counts detected in the hard band (2-10 keV) but only
7 in the soft (0.5-2 keV). The equivalent spectrum is inverted, with
$\Gamma=-1$, strongly implying heavy X-ray absorption of $\sim
10^{23}$~cm$^{-2}$. Despite this, it has the optical spectrum of a
type I AGN, a very broad H$\beta$ line (FWHM$\sim 9000$ km s$^{-1}$).
Clearly then, the optical broad line region of this object is not
obscured, while the X-rays are. It is at z=0.50 and the
2-10 keV luminosity is $\sim 10^{44}$~erg s$^{-1}$, classifying this
as a Seyfert 1, or low luminosity type I QSO. We stick
with the latter, ignoring the distinction between Seyferts
and QSOs, and furthermore adopt the traditional type I/II
classification based only on the optical/UV spectra.

{\bf AX J0836.2+5538 (S2):} This object is quite faint optically ($R=20.2$) but it shows clear evidence of a broad emission line
(8000 km s$^{-1}$ FWHM)  which we identify as Mg II at z=1.29. It is therefore a QSO with an X-ray luminosity of approximately $10^{45}$~erg s$^{-1}$. Given this is a type I AGN that is very bright in the X-ray band, perhaps the most interesting thing about this source is its relative optical faintness. 

{\bf AX J0836.6+5529 (S3):} We have discussed above the relative faintness of the Chandra sources in the error cirlce of this SHEEP source. The optical counterpart of the most likely Chandra object is the faintest of all we attempted spectroscopically (R=20.8). The signal-to-noise ratio of the spectrum is poor, but there does seem to be a broad emission line ($\sim 6000$ km s$^{-1}$), which we again tentatively identify as Mg {\sc ii}, this time at z=0.98. This object is therefore also most likely to be a standard QSO. 

{\bf AX J1035.1+3938 (S4):}
The hardness ratio of this object is similar to that of S1, 
again implying a heavy absorption. The optical
spectrum shows several narrow lines including H$\alpha$, H$\beta$ and
[O{\sc iii}] $\lambda$5007 at $z=0.11$.  It has an X-ray luminosity of
$L_{\rm X}$=$4 \times 10^{42}$~erg s$^{-1}$.  In order to classify
this object optically we have measured several line ratios, including
[O III] $\lambda$5007/H$\beta$ = 2, [N II] $\lambda$6583/H$\alpha$ =
0.5, and ([S II] $\lambda$6716 + $\lambda$6731)/H$\alpha$ = 0.5.
According to the diagnostic diagrams of Veilleux \& Osterbrock (1987)
this places the object roughly on the dividing line between AGN and
H{\sc ii} galaxies. Certainly, the spectrum is not that of a classical
Seyfert 2 and the object is likely to be undergoing a burst of star
formation which may dominate the optical light and excitation.

{\bf AX J1153.7+4619 (S5):}
This is the only object discussed in the current paper with a ROSAT counterpart, and therefore without a Chandra observation. The optical counterpart is again quite faint, with $R=20.6$, but shows strong, narrow O{\sc ii}, O{\sc iii} and H$\beta$ emission lines characteristic of a type II AGN at z=0.59. The inferred X-ray luminosty is $7 \times 10^{44}$~erg s$^{-1}$ in the 2-10 keV band making it a type II QSO. Most interestingly its  X-ray colors (from ASCA: paper I) show an effective $\Gamma\sim 1.7$, nowhere near as hard as those of the type I AGN S1 or the composite galaxy S4. This indicates within our sample an almost complete lack of correspondence between X-ray and optical measures of obscuration. 

\section{DISCUSSION}

X-ray surveys are the most efficient way of finding AGN, and hard
X-rays in particular have great importance as they are not
strongly biased against obscured objects unless the column densities
are huge. For example, in the SHEEP detection band of 5-10
keV a column density of $10^{23}$~cm$^{-2}$ (equivalent to an
extinction of $A_{\rm V} \sim 50$) only depresses the flux by $\sim
15\%$ at $z=0$, and even less at higher redshifts. Thus the content of
the survey should more closely represent the true AGN population
than, say, optical or soft X-ray surveys. Chandra
and XMM do the same, but for fainter sources. Our preliminary results
show that the SHEEP objects bear little resemblence to standard AGN
classes.

We find S1 to be very hard in the X-ray but shows broad
optical lines. Two similar objects were found in Paper 1, FBQS
J125829.6+35284 and CRSS J1429.7+4240  (see also
Comastri et al. 2001; Georgantopoulos et al. 2003), hinting that this
class of object - optical type 1s with heavily absorbed X-ray spectra
- may be very common. The implication is that the obscuring material
is dust free and/or very close to the nucleus. This is also the
situation in some bright local AGN (e.g. NGC 4151 - see paper I). Our
survey is showing that this phenomenon could be quite common in more
typical AGN making up the X-ray background, and which may evolve
differently to unobscured, broad line QSOs. This suggests that the
physical processes very close to the central black hole in galactic
nuclei evolve with cosmic time. This fascinating possibility clearly
deserves further investigation.

S2, also a broad line QSO, must also be obscured, based
both on its Chandra colour and the fact it was observed but not
detected by ROSAT. What is perhaps most interesting, however, is its
relative optical faintness. The detection of a broad line in
the rest-frame UV shows that the object cannot be heavily obscured in
the optical, and the object is therefore faint simply due to a high
\fxfopt\ ratio. Indeed, a QSO with the same \fxfopt\ ratio detected at
the flux limit of the Chandra 1Ms surveys would have R$>26$, making it
very difficult to identify spectroscopically. This shows the power of
the SHEEP survey in finding brighter objects which can be studied in 
more detail.  It also hints that many of the "optically
faint" sources found in the Chandra deep surveys (Mushotzky et
al. 2000; Alexander et al. 2001) may not have particularly
exotic properties.

This conclusion is strengthened when we consider 
S3. The optical spectrum is of poor quality, but if our identification is correct it is also a broad-line QSO.  Again one might classify this type of object as "optically faint" if it were detected in a deep survey.  The \fxfopt\ ratio based on the Chandra flux implies R=25 at the CDF-S flux limit. If the higher ASCA flux is more representative we would expect the object to appear at R$>28$ in a Chandra 1Ms survey. One must also ask why the ASCA and Chandra fluxes are so discrepant. Watanabe et al. (2002) concluded that large amplitude variability was at play when they observed similar behaviour in Chandra followup of two hard ASCA sources. To conclude that in our case would be premature, but we will be able to much more about this phenomenon when we have an analysis of the whole SHEEP sample. For example, if some SHEEP objects were spuriously detected (e.g. due to Eddington bias) we should observe the discrepancy only in the least significant sources, like S3. Otherwise the prospect of widespread transience or outburst in hard X-ray selected objects is most interesting, and would impact to a degree on the conclusions drawn from all flux-limited X-ray surveys. 

According to its optical spectrum, S4 is intermediate between an AGN and an H{\sc ii} galaxy. A  naive conversion from flux to luminosity would then make it the most X-ray luminous starburst known (c.f. Zezas, Georgantopolous \& Ward 1998; Moran et al. 1999). The very hard X-ray colours argue for substantial obscuration too, making the intrinsic luminosity even higher. It is perhaps more likely that the object hosts an obscured accreting black hole. 
Substantial obscuration can occur in the star-forming regions, blocking soft X-rays optical light from the nucleus.  The "intermediate" classification may also be in large part due to contamination of the AGN by the (extended) galaxy. Indeed, our photometry indicates that the whole galaxy is almost
2 magnitudes brighter than a 2" region centered on the nucleus (roughly where the spectrum was
obtained). 
A similar object observed at higher redshift would presumably therefore be completely dominated by stellar light. This dilution may go some way toward accounting for the fact that deeper surveys reveal many strong X-ray sources without AGN signatures (e.g. Mushotzky et al. 2000). 

Our final object, S5, is a candidate member of a long-sought after population: type II QSOs. These have been known to exist for quite some time (e.g. Hines \& Wills 1993) but excitement, and controversy, over their numbers have surrounded them due to their possible contribution to the X-ray background. They have been found in Chandra surveys (Stern et al. 2002; Norman et al. 2002), but they do not appear to be particularly common. On the other hand, Steidel et al. (2002) have shown that narrow-line objects are the dominant population of AGN in faint UV-selected galaxy samples at high redshift. S5 is also interesting because its X-ray colors from ASCA show relatively little obscuration: certainly less than the broad-line QSO S1
(see also Pappa et al. 2001). This emphasises an emerging phenomenon: that in hard X-ray surveys the X-ray and optical spectroscopic properties of AGN show almost no correspondence. This suggests it is time to revisit AGN classification schemes based on low redshift Seyferts: they are likely to tell us very little about black hole accretion at higher redshift. 

The SHEEP survey was designed to pick out bright examples of objects which make up the X-ray background. The survey seems to be fulfilling that promise, but the first indications are that these objects are quite unlike classical AGN. Given the relative lack of bias in hard X-ray selected samples, and their efficiency in finding accreting black holes, this indicates that much of what we think we know about what makes an AGN needs to be revised. Robust statistical conclusions await the results of the full survey, which should reveal much about the underlying AGN population. 

\section{Acknowledgements}

We are extremely grateful to Jules Halpern for the optical spectrum of AX J0140.1+0628, and thank the anonymous referee for a very helpful report. 
This research has made use of observations at  Kitt Peak National Observatory, National Optical Astronomy Observatory, which is operated by the Association of Universities for   Research in Astronomy, Inc. (AURA) under cooperative agreement with the National Science Foundation, of the NASA/IPAC Extragalactic database, which is operated by the Jet Propulsion Laboratory, Caltech, under contract with NASA, and of data obtained through the HEASARC online service, provided by NASA/GSFC. Skinakas Observatory is a collaborative project of the University of
Crete, the Foundation for Research and Technology-Hellas, and the
Max--Planck--Institut f\"ur extraterrestrische Physik. This work has been supported by a Chandra Guest Observer grant.


\bsp

\label{lastpage}
\end{document}